\documentclass[astrosymb,twocolumn]{aastex701}
\usepackage{amsmath}
\usepackage{graphicx}
\usepackage{rotating}

\received{January 23, 2026}
\revised{March 2, 2026}
\accepted{March 4, 2026}
\submitjournal{ApJ}

\begin{document}

\title{Predictions of Imminent Earth Impactors Discovered by LSST}

\author[orcid=0009-0005-9428-9590,sname='Chow']{Ian Chow}
\affiliation{DiRAC Institute and the Department of Astronomy, University of Washington, 3910 15th Avenue NE, Seattle, WA 98195, USA
}
\email[show]{chowian@uw.edu}
\correspondingauthor{Ian Chow}

\author[orcid=0000-0003-1996-9252,sname='Juric']{Mario Juri\'{c}}
\affiliation{DiRAC Institute and the Department of Astronomy, University of Washington, 3910 15th Avenue NE, Seattle, WA 98195, USA
}
\email{mjuric@astro.washington.edu}

\author[orcid=0000-0001-5916-0031,sname='Jones']{R. Lynne Jones}
\affiliation{Rubin Observatory, 950 N. Cherry Avenue, Tucson, AZ 85719, USA
}
\affiliation{Aston Carter, Suite 150, 4321 Still Creek Drive, Burnaby, BC V5C6S, Canada
}
\email{ljones.uw@gmail.com}

\author[orcid=0009-0006-2320-0306,sname='Kiker']{Kathleen Kiker}
\affiliation{Asteroid Institute, 20 Sunnyside Avenue, Suite 427, Mill Valley, CA 94941, USA
}
\email{kathleen@b612foundation.org}

\author[orcid=0000-0001-5820-3925,sname='Moeyens']{Joachim Moeyens}
\affiliation{DiRAC Institute and the Department of Astronomy, University of Washington, 3910 15th Avenue NE, Seattle, WA 98195, USA
}
\affiliation{Asteroid Institute, 20 Sunnyside Avenue, Suite 427, Mill Valley, CA 94941, USA
}
\email{moeyensj@uw.edu}

\author[orcid=0000-0001-6130-7039,sname='Brown']{Peter G. Brown}
\affiliation{Department of Physics and Astronomy, University of Western Ontario, 1151 Richmond St, London, ON N6A 3K7, Canada
}
\affiliation{Western Institute for Earth and Space Exploration, University of Western Ontario, Perth Drive, London, ON N6A 5B7, Canada
}
\email{pbrown@uwo.ca}

\author[orcid=0000-0003-3313-4921,sname='Heinze']{Aren N. Heinze}
\affiliation{DiRAC Institute and the Department of Astronomy, University of Washington, 3910 15th Avenue NE, Seattle, WA 98195, USA
}
\email{aheinze@uw.edu}

\author[0009-0005-5452-0671,sname='Kurlander']{Jacob A. Kurlander}
\affiliation{DiRAC Institute and the Department of Astronomy, University of Washington, 3910 15th Avenue NE, Seattle, WA 98195, USA
}
\email{jkurla@uw.edu}

\begin{abstract}
Imminent impactors are natural bodies discovered in space before impacting the Earth. They provide a rare opportunity to characterize individual near-Earth objects (NEOs) in great detail as asteroids in space, meteors in Earth's atmosphere and meteorites on the ground.
The Vera C. Rubin Observatory's upcoming Legacy Survey of Space and Time (LSST) is expected to transform our understanding of the NEO population.
In this work, we evaluate LSST's expected discovery performance for imminent impactors using $343$ meter-size objects previously recorded in NASA's CNEOS database as fireballs impacting Earth's atmosphere.
We simulate pre-impact observations of these CNEOS impactors with the \texttt{Sorcha} survey simulator under LSST's default three-night discovery strategy and a one-night strategy for fast-moving objects that relies on matching aligned streaks in two exposures on the same night.
We estimate that LSST will discover $\sim1-2$ meter-size and larger imminent impactors per year, representing $\sim4\%$ of all Earth impactors $\gtrsim1$ m in diameter and almost doubling the current discovery rate of imminent impactors. The median time of discovery and median time of first observation for impactors discovered in our simulations are $\sim1.57$ and $\sim3.06$ days before impact, respectively.
The spatial distribution of the 11 previously discovered imminent impactors is biased towards the Northern Hemisphere, where the observatories that discovered them are located. We find a similar trend towards Southern Hemisphere impacts in our simulated LSST detections of the CNEOS impactors, suggesting
Rubin will provide a powerful counterpart to existing asteroid surveys primarily located in the Northern Hemisphere.

\end{abstract}




\section{Introduction}\label{sec:introduction}

Asteroids and comets are primordial remnants of the processes that shaped the early Solar System, recording a wealth of information about the Solar System's formation and early history in their physical and orbital properties. 
An estimated $1$ billion of these bodies are near-Earth objects (NEOs; objects with perihelion distance $q < 1.3$ AU) $1$ meter or larger in diameter \citep[][]{harris_population_2015, harris_population_2021}, of which $\sim20\%$ pass close enough to the Earth such that orbital perturbations occurring over timescales of a century can dynamically evolve them onto Earth-crossing orbits \citep[e.g.][]{jones_large_2018}.
These objects occasionally collide with the Earth, allowing them to be observed as bright meteors (termed fireballs or bolides) in Earth's atmosphere. In some cases, parts of them can reach the ground to be recovered as meteorite falls, of which $\sim1-20$ m asteroids are the primary source \citep[][]{borovicka_are_2015}.
Linking recovered meteorites to observed meteors is now done frequently, especially with the deployment of dedicated, modern meteor observing networks including the European Fireball Network \citep[EN;][]{spurny_automation_2006, borovicka_data_2022-1}, Fireball Recovery and InterPlanetary Observation Network \citep[FRIPON;][]{colas_fripon_2020}, Global Fireball Observatory \citep[GFO;][]{devillepoix_global_2020} and Global Meteor Network \citep[GMN;][]{vida_global_2021}.
At least $75$ meteorite falls have been observed as meteors \citep{jenniskens_review_2025}, allowing for numerous analyses linking meteorite classes to asteroid families and source regions of the main asteroid belt (see \citet{binzel_near-earth_2015} and \citet{jenniskens_review_2025} for reviews).

However, discovering an asteroid in space before Earth impact is more difficult and requires a significant amount of luck; only eleven such objects have been found to date (listed in Table \ref{tab:imminent_impactors}) and in all cases less than a day before impact.
Throughout this work, we use the term ``imminent impactors" to refer specifically to such objects discovered before Earth impact and the term ``Earth impactors" or simply ``impactors" to refer to natural bodies that impact the Earth in general.

While rare, imminent impactors represent a unique opportunity for interdisciplinary studies combining measurements of these objects in three different regimes: as asteroids in space, meteors in Earth's atmosphere, and recovered meteorites on the ground, furthering a holistic understanding of them. For example, a comparison of telescopic reflectance spectra for the imminent impactor 2008 TC3 with spectra of meteorites recovered after impact revealed a link between F/B-type asteroids and ureilites, a type of stony achondrite \citep{jenniskens_impact_2009, bischoff_asteroid_2010}.
\begin{table*}
    \centering
    \begin{tabular}{l|l|l|l|l|l}
    \hline
        Object & UTC Impact Date/Time & Discoverer & Time of Discovery & Diameter & Reference  \\
        & (YYYY-MM-DD HH:MM:SS) & & (days before impact) & (m) & \\ \hline
        2008 TC3 & 2008-10-07 02:45:40 & Mt. Lemmon Survey & $0.837$ & $4.1$
        & \citet{jenniskens_impact_2009} \\
        2014 AA & 2014-01-02 03:04:50 $\pm$ $373$ s & Mt. Lemmon Survey & $0.866$ & $2.1$* & \citet{farnocchia_trajectory_2016} \\ 
        2018 LA & 2018-06-02 16:44:12 & Mt. Lemmon Survey & $0.354$ & $1.56$ & \citet{jenniskens_impact_2021} \\
        2019 MO & 2019-06-22 21:25:47 & ATLAS-MLO & $0.483$ & $4.7$* &
        Minor Planet Center\footnote{\href{https://minorplanetcenter.net/mpec/K19/K19M72.html}{minorplanetcenter.net/mpec/K19/K19M72.html}} \\
        2022 EB5 & 2022-03-11 21:22:45 & Konkoly Observatory & $0.0819$ & $5-6$ & \citet{geng_near-earth_2023} \\
        2022 WJ1 & 2022-11-19 08:26:40 & Mt. Lemmon Survey & $0.148$ & $0.50$
        & \citet{kareta_telescope--fireball_2024} \\
        2023 CX1 & 2023-02-13 02:59:13 & Konkoly Observatory & $0.279$ & $0.72$
        & \citet{egal_catastrophic_2025} \\
        2024 BX1 & 2024-01-21 00:32:38 & Konkoly Observatory & $0.115$ & $0.44$ & \citet{spurny_atmospheric_2024} \\
        2024 RW1 & 2024-09-04 16:39:32 & Mt. Lemmon Survey & $0.455$ & $3.3$ 
        & \citet{ingebretsen_apache_2025} \\
        2024 UQ & 2024-10-22 10:54:48 & ATLAS-HKO & $0.0736$ & $0.92$* & Minor Planet Center\footnote{\href{https://minorplanetcenter.net/mpec/K24/K24U49.html}{minorplanetcenter.net/mpec/K24/K24U49.html}} \\
        2024 XA1 & 2024-12-03 16:14:52 & Kitt Peak & $0.431$ & $1$ & \citet{gianotto_fall_2025} \\
    \end{tabular}
    \caption{All $11$ imminent impactors (i.e. Earth-impacting asteroids discovered in space before impact) as of 2026 January 1, ordered by their impact date.
    }
    \raggedright
    \textbf{Note:} For objects whose size is not explicitly given by the provided reference (marked with an asterisk), it is estimated from the $H$-magnitude provided by the 
    Minor Planet Center
    assuming a reference geometric albedo of $p_v = 0.18$ appropriate for small NEOs \citep[$H < 22$;][]{nesvorny_neomod_2024}. The large uncertainty in the impact time of 2014 AA is because the impact was not directly observed but was inferred from infrasound measurements \citep{farnocchia_trajectory_2016}.
    \label{tab:imminent_impactors}
\end{table*}



Imminent impactors are also a scientifically interesting population as they are among the smallest objects ever characterized telescopically. For example, the imminent impactor 2024 BX1 is both the smallest and fastest-rotating natural body ever observed in space, with a diameter of 
$\sim44$ cm \citep{spurny_atmospheric_2024, bischoff_cosmic_2024}
and rotation period of $2.59$ s \citep{devogele_aperture_2024}.
The rotation periods measured for other imminent impactors are similarly fast, typically on the order of minutes \citep{jenniskens_impact_2009, jenniskens_impact_2021, ingebretsen_apache_2025}.

The physical structure of large meteoroids and very small asteroids is believed to be qualitatively different from larger asteroids, based on studies of their atmospheric fragmentation from meteor observations and their rotation rates in space. While most asteroids larger than $\sim200$ m are strengthless, gravitationally bound ``rubble piles" \citep[e.g.][]{pravec_fast_2000}, decimeter- to decameter-size bodies are believed to be heterogeneous objects composed of reassembled collisional debris weakly cemented by impact melting onto a stronger core of macroscopically cracked monolithic material with some inherent tensile strength \citep[][]{borovicka_two_2020, chow_decameter-sized_2026}.

Telescopic observations of imminent impactors can provide unique information on the composition of very small asteroids, probing a size regime typically limited to meteor and meteorite studies. For example, the telescopic reflectance spectrum of the imminent impactor 2022 WJ1 unusually suggested that it had very little regolith on its surface, possibly due to rapid rotation \citep{kareta_telescope--fireball_2024}.

Many of these imminent impactors were found by large-scale asteroid surveys, including the Catalina Sky Survey \citep[CSS;][]{larson_catalina_1998, christensen_catalina_2012} and the Asteroid Terrestrial-impact Last Alert System \citep[ATLAS;][]{tonry_atlas_2018}.
Upcoming space- and ground-based programs such as NEO Surveyor \citep{mainzer_near-earth_2023} and the Vera C. Rubin Observatory \citep[Rubin;][]{ivezic_lsst_2019} are expected to transform our understanding of the NEO population in general and imminent impactors in particular. This work focuses on the latter.

The Vera C. Rubin Observatory \citep{ivezic_lsst_2019} completed construction and commissioning in 2025 October. Rubin is the National Science Foundation's (NSF) new flagship astronomical survey facility, housing the $8$m-class Simonyi Survey Telescope and the $3.2$ Gigapixel LSSTCam camera. Together with a petascale automated data analysis system, Rubin is an integrated, dedicated survey facility whose goal is to collect and process the data for the Legacy Survey of Space and Time (LSST), a dataset covering 24,000 deg$^2$ of the sky (18,000 to uniform depth) to single-exposure magnitude of $m_r \sim 24.0$ in \textit{ugrizy} bands. To build the LSST dataset, Rubin will repeatedly observe the sky, typically collecting pairs of observations of each pointing 30 minutes apart in a night, and returning to the same pointing every 3-4 nights. This cadence will result in $\sim 800$ re-observations of the same area over the survey's 10-year duration, providing a time-domain dataset of broad utility for science cases ranging from understanding the nature of dark energy to taking a census of the Solar System.

Once LSST data collection begins in early 2026, it will deliver an unprecedented, comprehensive census of small bodies of the Solar System. \cite{kurlander_predictions_2025} show that LSST is expected to identify approximately $3.9$ million previously unknown Solar System objects, representing increases of $\sim3.5$x for main-belt asteroids (MBAs) and $\sim10$x for trans-Neptunian objects (TNOs), with other populations falling in between. This sample will make it possible to study the detailed structure and evolution of nearly all small-body populations, from the outer reaches of the Solar System to objects that approach the Earth.

\citet{kurlander_predictions_2025} also predict LSST to increase the number of known NEOs to $127,000$, $\sim5$x larger than the currently known population. This estimate is derived by directly simulating Rubin observations of a synthetic NEO population drawn from \texttt{NEOMOD3} \citep{nesvorny_neomod_2024}, a debiased NEO orbital distribution model.
However, \citet{kurlander_predictions_2025} do not specifically investigate the Earth impactor population and undersample the population of $1-10$ m-size NEOs (the size range of most known imminent impactors) by a factor of four.
Moreover, \citet{chow_decameter-sized_2025} have previously identified an order-of-magnitude discrepancy between the observed Earth impact rate of $\sim10$-m size objects (the smallest size at which \texttt{NEOMOD3} is calibrated to real data) based on fireball data and the expected impact rate inferred from \texttt{NEOMOD3}, suggesting that the population of Earth impactors at these sizes may not be accurately represented by the model.



In this work, we develop a complementary technique to that of \citet{kurlander_predictions_2025} to extend these simulations into the meter-size regime and provide estimates of LSST's absolute discovery yield and completeness for Earth impactors. Rather than starting with a synthetic NEO population model as \citet{kurlander_predictions_2025} do, we instead use an empirically calibrated sample of actual meter-size and larger Earth impactors recorded by satellite-based United States Government (USG) sensors since 1994 as fireballs in Earth's atmosphere. The USG dataset is believed to represent a nearly complete census of actual Earth impactors in the multi-meter size range, allowing us to treat it as a stand-in population for future impactors\footnote{Any incompleteness in the sample results in our estimates being lower bounds, as we discuss later.}. We run this population through \texttt{Sorcha} \citep{merritt_sorcha_2025, holman_sorcha_2025}, a new high-fidelity Solar System survey simulator, to see which of these impactors could have been detected by LSST if the survey had been running at the time of impact. Finally, by applying plausible tracklet identification and linking strategies, we derive estimates for how many historical impactors Rubin could have been alerted to before impact. This approach sets strong, model independent lower limits on what to expect from LSST in the years to come.

A companion paper to this work by \citet[][in review]{cheng_assessing_2026} analyzes LSST's relative discovery efficiency for a synthetic population of Earth impactors larger than those considered here ($\gtrsim10$ m in diameter) generated using \texttt{NEOMOD3}. 

This paper is structured as follows: Section \ref{sec:data} describes the fireball data used in this work. In Section \ref{sec:visibility}, we compute the brightness of each of the $343$ recorded impactors during their final Earth approach in space to determine when they would become detectable by LSST and to demonstrate the feasibility of our study using a simplified approach. In Section \ref{sec:simulated_discoveries}, we use \texttt{Sorcha}, along with a version of LSST's baseline cadence modified to run from 1994 to the present day, to precisely simulate LSST observations of the impactors as if the survey had been running continuously throughout that time. Section \ref{sec:yield_completness} provides estimates of the discovery rate and completeness for Earth impactors based on the results of the \texttt{Sorcha} simulations. We discuss our results and future opportunities for imminent impactor science that LSST could enable in Section \ref{sec:discussion}.
Finally, Section \ref{sec:conclusions} summarizes the key conclusions of our study.


\section{CNEOS Fireball Data} \label{sec:data}

The data used in this work are primarily obtained from NASA's Center for Near Earth Object Studies (CNEOS) Fireball and Bolide Database\footnote{\href{https://cneos.jpl.nasa.gov/fireballs/}{cneos.jpl.nasa.gov/fireballs/}}, which has recorded more than $1000$ fireball events observed globally by USG satellites in Earth orbit \citep{gehrels_detection_1995} over several decades.
This database contains the largest and most comprehensive set of meter-size and larger Earth impactors to date, with near-global coverage \citep[$\sim80\%$ of Earth's surface;][]{brown_flux_2002}.

We begin by filtering the CNEOS database (as of 2026 January 1) to obtain a set of $343$ Earth impactors with known impact energies and state vectors at the time of impact and which do not originate on bound geocentric orbits (i.e. with geocentric impact velocity $V_g > 11.2$ km/s), spanning the time period from 1994 February 1 to 2026 January 1. This velocity filtering removes $4$ impacts from the initial dataset.

This set includes five impactors whose USG sensor-recorded state vectors are not reported on the CNEOS website but have been published: the 1994 February 1 Marshall Islands \citep{tagliaferri_analysis_1995, mccord_detection_1995}, 1994 June 15 St. Robert \citep{brown_fall_1996}, 2000 January 18 Tagish Lake \citep{brown_entry_2002}, 2003 March 27 Park Forest \citep{brown_orbit_2004} and 2004 September 3 Antarctica \citep{klekociuk_meteoritic_2005} fireballs.
We also reverse the $z$-component of 2008 TC3's CNEOS-reported velocity vector from $3.8$ km/s to $-3.8$ km/s following \citet{pena-asensio_orbital_2022}, who identified it as a likely typographical error on the CNEOS website.

The fireballs recorded in the CNEOS database do not have associated uncertainties and so the accuracy of the state vectors on a per-object basis is unknown.
However, the CNEOS-reported impact energies are generally consistent with those determined by ground-based optical \citep{devillepoix_observation_2019} and infrasound \citep{gi_refinement_2017} measurements, as well as by the satellite-based Geostationary Lightning Mapper \citep[GLM;][]{wisniewski_determining_2024}. The locations have also been found to be similarly accurate when compared to ground-based observations \citep{devillepoix_observation_2019, brown_proposed_2023}.
The entry velocities and radiants are less accurately measured, and many previous works have attempted to quantify their uncertainties \citep[e.g.][]{devillepoix_observation_2019, brown_proposed_2023, pena-asensio_oort_2024, hajdukova_no_2024, chow_decameter-sized_2025, pena-asensio_error_2025}. We discuss these uncertainties further in Section \ref{sec:visibility}.

The diameter of each object is computed using its reported velocity and total impact energy, assuming a spherical shape and bulk density of $1500$ kg/m$^3$, based on other bulk density estimates of 
small
NEOs \citep[see e.g. Section $2$ of][and references therein]{chow_decameter-sized_2025}. 
While it is difficult to quantify the uncertainty in bulk density, it is unlikely to have a significant effect on our results;
since diameter is proportional to the negative one-third power of bulk density, even the theoretical maximum bulk density of $3500$ kg/m$^3$ (corresponding to zero porosity silicate rock) represents a less than $25\%$ decrease in diameter compared to our nominal bulk density of $1500$ kg/m$^3$, or a decrease in brightness of $\sim0.6$ mag over the size range considered in this work.

\begin{figure*}
    \centering
    \includegraphics[width=1.\linewidth]{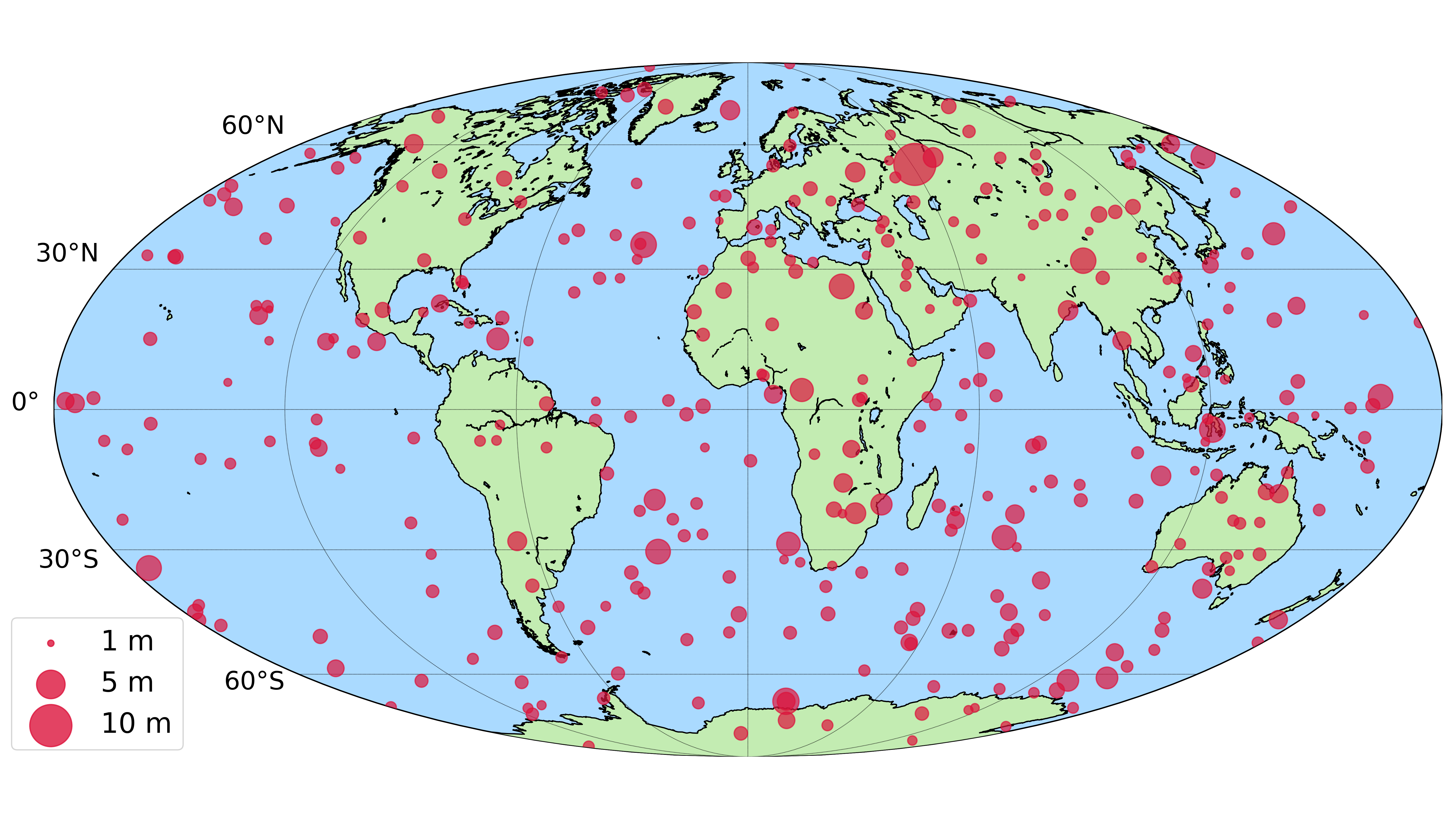}
    \caption{An equal-area Mollweide projection map showing the impact locations (red dots) of all $343$ Earth impactors with known state vectors and impact energies recorded in NASA's Center for Near-Earth Object Studies (CNEOS) Fireball and Bolide Database from 1994 February 1 to 2026 January 1. The estimated diameter of each object (ranging from $\sim0.75-24$ m) is indicated by the relative size of its dot. The spatial distribution of impacts over Earth's surface appears generally uniform.}
    \label{fig:cneos_map}
\end{figure*}

The computed sizes range from $\sim0.75-24$ m in diameter, with the upper bound being the size of the 2013 Chelyabinsk asteroid 
\citep[estimated as $\sim19$ m by][using a higher bulk density]{borovicka_trajectory_2013}, 
the largest Earth impactor for which observational data is available. Figure \ref{fig:cneos_map} shows a map of impact locations of the $343$ impactors analyzed in this study, with each object's estimated size indicated by its relative size on the map.

Each impactor's equivalent absolute $V$-band magnitude, $H$, is then estimated from its diameter, $D$, in kilometers, and its geometric $V$-band albedo, $p_v$, using the expression \citep[][]{tedesco_iras_1992, harris_revision_1997}
\begin{equation}
    H = 5\left(3.1236 - \log_{10}\left(D\sqrt{p_v}\right)\right). \label{eqn:H_diameter}
\end{equation}
The visible albedo distribution of known asteroids is bimodal \citep[e.g.][]{murray_using_2023, ge_asteroid_2025} with peaks roughly corresponding to the median albedos for S- and C-type asteroids \citep[$p_v \approx 0.25$ for S-types and $p_v \approx 0.06$ for C-types;][]{mainzer_neowise_2011, demeo_taxonomic_2013}, which are the most common NEO types in approximately equal proportions \citep[][]{marsset_debiased_2022}. They also dominate the meteorite flux on Earth, with about half of the total coming from S-types and the other half from C-types \citep{broz_young_2024, broz_source_2024}.

Here we randomly assume each impactor to be either an S- or C-type asteroid with equal probability and use the corresponding median albedo as a reference value to convert from size to $H$-magnitude. We note that while the albedo distribution of larger NEOs is more complex than assumed here and depends on asteroid size and source region \citep[e.g.][]{wright_albedo_2016, binzel_compositional_2019, marsset_debiased_2022, nesvorny_neomod_2024}, there are few constraints for NEO albedos and colors at meter sizes. As such, we choose to adopt a relatively simple assumption in this work that broadly captures the two albedo modes for S- and C-type asteroids, which dominate the meter-size Earth impactor population.

This assumption is also in reasonable quantitative agreement with the four known imminent impactors (2008 TC3, 2022 EB5, 2022 WJ1 and 2023 CX1) whose albedos have been independently constrained from recovered meteorites and/or using mass estimates derived from ablation modelling of their fireball light curves; 2022 WJ1 and 2023 CX1 have typical S-type albedos of $0.27$ and $0.28$ respectively \citep{kareta_telescope--fireball_2024, egal_catastrophic_2025}, while 2008 TC3 and 2022 EB5 have very low C-type albedos of $0.046$ and $0.025$ \citep{jenniskens_impact_2009, geng_near-earth_2023}.

\begin{table*}
    \centering
    \begin{tabular}{l|l|l|l|l}
         Object & Telescopic $H$ & CNEOS $H$ & CNEOS $H$ & Reference \\
          &  & (S-type) & (C-type) & \\
         \hline
         2008 TC3 & $30.86$ & $29.16$ & $30.71$ 
         & \citet{kozubal_photometric_2011} \\
         2018 LA & $30.55$ & $29.57$ & $31.05$ 
         & Minor Planet Center \\
         2019 MO & $29.12$ & $28.02$ & $29.50$ 
         & Minor Planet Center \\
         2022 EB5 & $31.33$ & $28.57$ & $30.12$ & Minor Planet Center \\
         2024 RW1 & $30.92$ & $30.89$ & $32.44$ & \citet{ingebretsen_apache_2025} \\ \hline
    \end{tabular}
    \caption{Comparison of estimated $H$-magnitudes from telescopic observations and from CNEOS impact energy/velocity using a reference geometric albedo of $p_v \approx 0.25$ appropriate for S-type asteroids and $p_v \approx 0.06$ appropriate for C-type asteroids \citep{mainzer_neowise_2011, demeo_taxonomic_2013} for the subset of CNEOS impactors also discovered telescopically prior to impact.}
    \label{tab:h_mags}
\end{table*}

We note that five of the eleven known imminent impactors (2008 TC3, 2018 LA, 2019 MO, 2022 EB5 and 2024 RW1) have CNEOS-recorded state vectors and impact energies.
As an overall test of our procedure in this section, Table \ref{tab:h_mags} compares the estimated $H$-magnitudes of these objects from CNEOS impact energies and velocities (for S- and C-type albedos) to those reported by the
Minor Planet Center (MPC),
or from more detailed photometry in the literature when available. 

The telescopically determined $H$-magnitudes for 2008 TC3 and 2024 RW1 are close to the CNEOS estimates for the S- and C-type albedo modes, with 2018 LA and 2019 MO falling in between.
Several previous works have independently confirmed a bias in the MPC's reported asteroid $H$-magnitudes as being systematically brighter than the actual $H$-magnitudes by $\sim0.4$ mag \citep[][]{juric_comparison_2002, pravec_absolute_2012, veres_absolute_2015}, which could shift 2018 LA and 2019 MO closer to the C-type mode.
The large discrepancy for 2022 EB5 is likely due to its extremely low albedo of $0.011-0.025$, as mentioned earlier. Using this albedo range instead yields a CNEOS-derived $H$-magnitude of $31.07-31.96$, which is consistent with the MPC-reported value.

Considering all of these factors, our absolute magnitude estimates -- while having some variance from albedo and to a lesser extent bulk density -- are likely systematically correct to within a few tenths of a magnitude.

In all, our analysis leaves us with an input dataset of $343$ real meter-size Earth impactors spanning the time period from 1994 January 1 to 2026 January 1, which we use as a representative impactor population for the LSST impactor discovery yield experiments conducted in this work. We make the data publicly available at \url{https://dirac.us/qkw}.





\section{Impactor detectability at LSST depths: a feasibility check} \label{sec:visibility}

We begin by considering a simplified toy model: how soon before impact could each of the CNEOS impactors be detected assuming all-sky coverage to $m_r = 24.0$ (a reasonable choice for fiducial $r$-band LSST single-exposure depth)? We determine this by integrating all impactors backwards and observing at what point they reach the $m_r = 24.0$ detectability threshold.

The state vector of each impactor is recorded by CNEOS as a position defined by the geodetic latitude, longitude and elevation, and a Cartesian $x$,$y$,$z$ velocity vector defined in the Earth-centered, Earth-fixed (ECEF) equatorial coordinate frame, at the time of impact.
We first convert each impactor's ECEF state vector to Solar System barycentric equatorial coordinates using the
Jet Propulsion Laboratory's (JPL)
DE441 ephemerides and the time of impact. Next, we integrate each object backwards in time $14$ days before impact and record its position at one-hour intervals.
The integrations for all simulations presented in this paper are conducted using \texttt{ASSIST} \citep{holman_assist_2023}, a software package for generating ephemeris-quality integrations of Solar System test particles. 
\texttt{ASSIST} is an extension of the \texttt{REBOUND} software \citep{rein_rebound_2012} for numerical $N$-body integration and uses the IAS15 integrator \citep{rein_ias15_2015} to integrate test particle trajectories in the field of the Sun, Moon, planets, and $16$ massive asteroids (with positions from the DE441 ephemerides), including the most significant gravitational harmonics of Earth, general relativistic and non-gravitational effects.
For all simulations in this work, we use accuracy parameter $\epsilon = 10^{-6}$, initial timestep of $10^{-6}$, minimum timestep of $10^{-9}$ (all values in units of days), and the timestep criterion of \citet{rein_ias15_2015} for the IAS15 integrator. These values were empirically determined using a grid search over parameter space to balance accuracy and compute time when integrating close encounters of Earth impactors. For all other parameters, we use the default values of \texttt{ASSIST} and \texttt{REBOUND}.


The apparent $V$-band magnitude of the object, $V_\mathrm{app}$, is then estimated at each step using the IAU standard $H$-$G$ asteroid magnitude system \citep{bowell_application_1989} as
\begin{align}
    V_\mathrm{app} = H  & - 2.5\log_{10}\left(\left(1 - G\right)\phi_1\left(\alpha\right) + G\phi_2\left(\alpha\right)\right) \nonumber \\ 
    & + 5\log_{10}\left(r\Delta\right)\text{,} \label{eqn:hg_mag}
\end{align}
where $H$ is the absolute $V$-band magnitude computed in Section \ref{sec:data}, $G$ is the slope parameter (here assumed to be $G = 0.15$ for all objects), $r$ and $\Delta$ are the asteroid's heliocentric and geocentric distance respectively, and the phase functions $\phi_1\left(\alpha\right)$, $\phi_2\left(\alpha\right)$ are defined as
\begin{align}
    \phi_1\left(\alpha\right) &= \exp\left(-3.33\left(\tan\frac{\alpha}{2}\right)^{0.63}\right) \nonumber \\ 
    \phi_2\left(\alpha\right) &= \exp\left(-1.87\left(\tan\frac{\alpha}{2}\right)^{1.22}\right)\text{,} \label{eqn:phis}
\end{align}
where $\alpha$ is the phase angle of the asteroid with respect to the Sun. Finally, each object's apparent $V$-band magnitude is converted to LSST $r$-band using the appropriate transformations provided by the Rubin simulation suite for its taxonomic class assumed earlier\footnote{The colors are provided in a Jupyter notebook at \url{https://github.com/lsst/rubin_sim_notebooks/blob/d7c3321161de9451288fb9661ee83cb189801753/photometry/solar_system_object_colors.ipynb}}.


\begin{figure*}
    \centering
    \includegraphics[width=\linewidth]{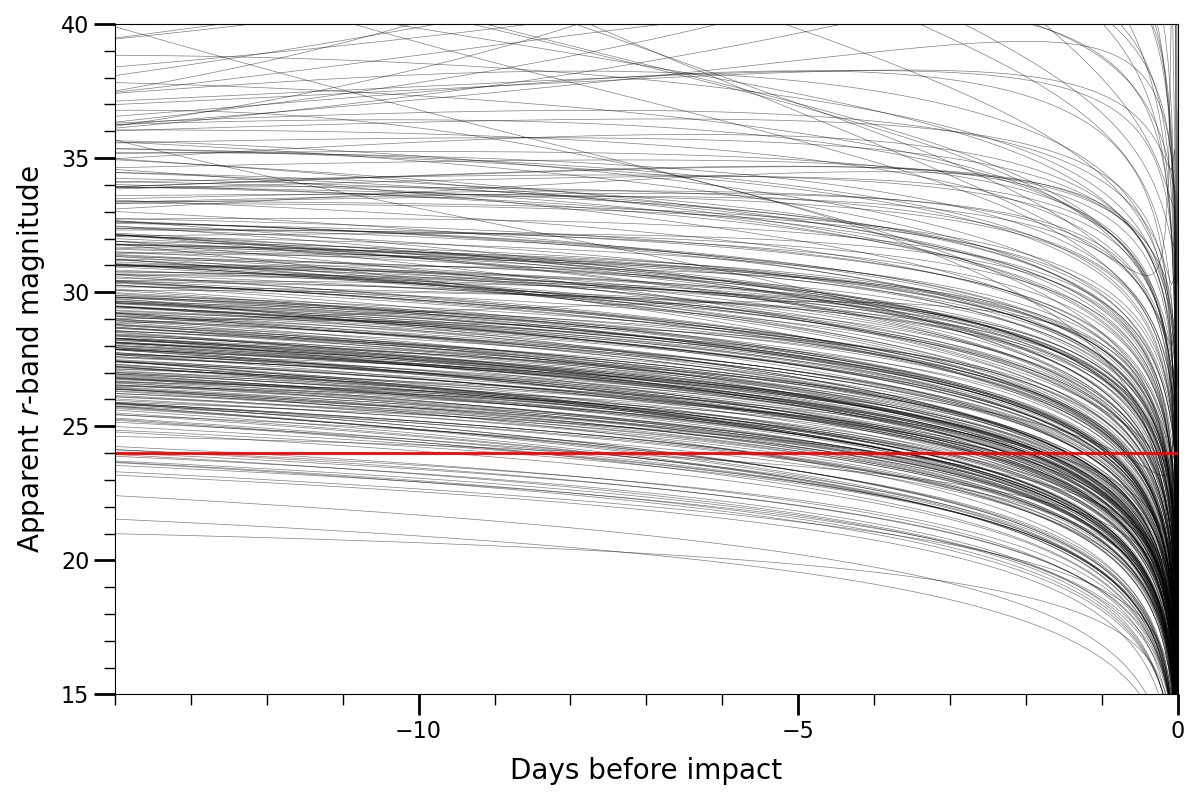}
    \caption{The apparent LSST $r$-band magnitude of $343$ CNEOS-recorded Earth impactors (black solid lines) at one-hour intervals during the $14$ days prior to impact. LSST's approximate single-image $r$-band depth of $m_r \sim 24.0$ is indicated by the red solid line. Most impactors would remain too faint to be detectable by LSST on their final approach until a few days before impact. However, nearly all of them ultimately cross the fiducial magnitude threshold, bringing them into the range of LSST visibility (see Figure~\ref{fig:cumulative_visible}).
    }
    \label{fig:r_mags}
\end{figure*}

Figure \ref{fig:r_mags} shows the resulting apparent $r$-band magnitude for each of the $343$ CNEOS impactors during the $14$ days before impact, with LSST's fiducial single-exposure depth of $m_r = 24.0$ indicated on the plot as a solid red horizontal line. As expected, the majority of objects exhibit a steep monotonic brightness drop on their final approach to the Earth due to the rapidly decreasing observer-object distance, with a few having more complex light curves due to the rapidly changing phase on final approach. This is especially noticeable for objects approaching from the day side as their magnitude can drop as they come between the observer and the Sun, even though the geometric distance becomes smaller. 
\begin{table*}
    \centering
    \begin{tabular}{c|c|c}
         Parameter & High-$D_D$ Median Error ($\pm 1\sigma$) & Low-$D_D$ Median Error ($\pm 1\sigma$) \\
         & ($\text{Yr.}<2018$ \& $E_i < 0.45$ kT TNT) & ($\text{Yr.}\geq2018$ \textbar $E_i \geq 0.45$ kT TNT) \\
         \hline
         $\mathbf{V}$ (km/s) &  $6.05_{-2.21}^{+1.44}$ & $0.55_{-0.45}^{+0.37}$ \\
         Az. ($^\circ$) & $17.32_{-4.24}^{+37.20}$ & $2.12_{-1.83}^{+2.61}$ \\ 
         Alt. ($^\circ$) & $13.46_{-12.38}^{+14.21}$ & $2.52_{-1.88}^{+1.46}$ \\ \hline 
    \end{tabular}
    \caption{Estimated median errors and $1\sigma$ upper and lower standard deviations in geocentric speed (impact speed relative to Earth prior to gravitational acceleration), local azimuth and altitude (elevation) angle for two groups of calibrated CNEOS fireballs from \citet{pena-asensio_error_2025}.}
    \raggedright
    \textbf{Note}: The calibrated fireballs are classified by \citet{pena-asensio_error_2025} into two groups based on the $D_D$ orbital dissimilarity criterion of \citet{drummond_test_1981} when comparing to ground-based observations. The high-dissimilarity group is composed of those fireballs recorded before 2018 and with impact energies lower than $0.45$ kilotons of TNT ($1$ kT TNT $=4.184 \times 10^{12}$ J). The low-dissimilarity group is composed of all other fireballs. The entry azimuth and altitude uncertainties were provided by E. Pe\~{n}a-Asensio, personal communication (2025).
    \label{tab:cneos_uncertainties}
\end{table*}

Figure \ref{fig:cumulative_visible} shows the cumulative fraction of CNEOS impactors that would be observable by LSST as a function of time before impact, both for the $343$ nominal objects as well as clones we generate to assess the effect of orbital uncertainties (discussed in Section \ref{sec:uncertainty}). We estimate that $10\%$ of CNEOS impactors are detectable by LSST at least $\sim6.5$ days before impact, and that $50\%$ are detectable at least $\sim1$ day before impact. This bolsters the overall feasibility of Rubin as an imminent impactor detection machine, motivating the effort to perform the detailed simulations of Section~\ref{sec:simulated_discoveries}.

\subsection{Impactor Orbital Uncertainty}\label{sec:uncertainty}

We now consider whether uncertainties in the impactor orbits could materially change Figure~\ref{fig:r_mags}.
While the velocity vectors reported by CNEOS do not have associated uncertainties, \citet{pena-asensio_error_2025} have empirically estimated CNEOS orbital uncertainties using a set of $18$ calibrated fireballs from the CNEOS database that also have independent ground-based measurements. Using the $D_D$ orbital dissimilarity criterion of \citet{drummond_test_1981}, \citet{pena-asensio_error_2025} divide these calibrated fireballs into two discrete groups with high and low orbital dissimilarities when comparing CNEOS-derived orbits to those derived from ground-based observations, with the high-$D_D$ group composed of those fireballs recorded before $2018$ and with impact energies lower than $0.45$ kilotons of TNT ($1$ kT TNT $=4.184 \times 10^{12}$ J) and the low-$D_D$ group composed of all other fireballs. \citet{pena-asensio_error_2025} find larger orbital uncertainties in the high-dissimilarity group, suggesting that the decreased uncertainty for less energetic impacts starting at the beginning of $2018$ is likely due to the deployment of new USG sensors.

We use the estimated CNEOS uncertainties of \citet{pena-asensio_error_2025}, summarized in Table \ref{tab:cneos_uncertainties}, to evaluate the robustness of our pre-impact visibility analysis against measurement error.
We generate $1000$ clones of each impactor by drawing Monte Carlo samples from the speed, azimuth and altitude uncertainty distributions in Table \ref{tab:cneos_uncertainties} (assuming they are uncorrelated) based on the impact date and energy, and adding the sampled uncertainties to the impactor's state vector. Clones with local altitude angle less than $0^\circ$ or greater than $90^\circ$ or that originate on bound geocentric orbits are discarded as unphysical. Each clone is then integrated backwards in time and its apparent $r$-band magnitude over the $14$ days before impact is computed using the same procedure as for the nominal impactors. Finally, we record the earliest time at which each impactor (both nominal and cloned) becomes detectable at LSST's depth (i.e. when it becomes brighter than $m_r = 24.0$). 

\begin{figure*}
    \centering
    \includegraphics[width=\linewidth]{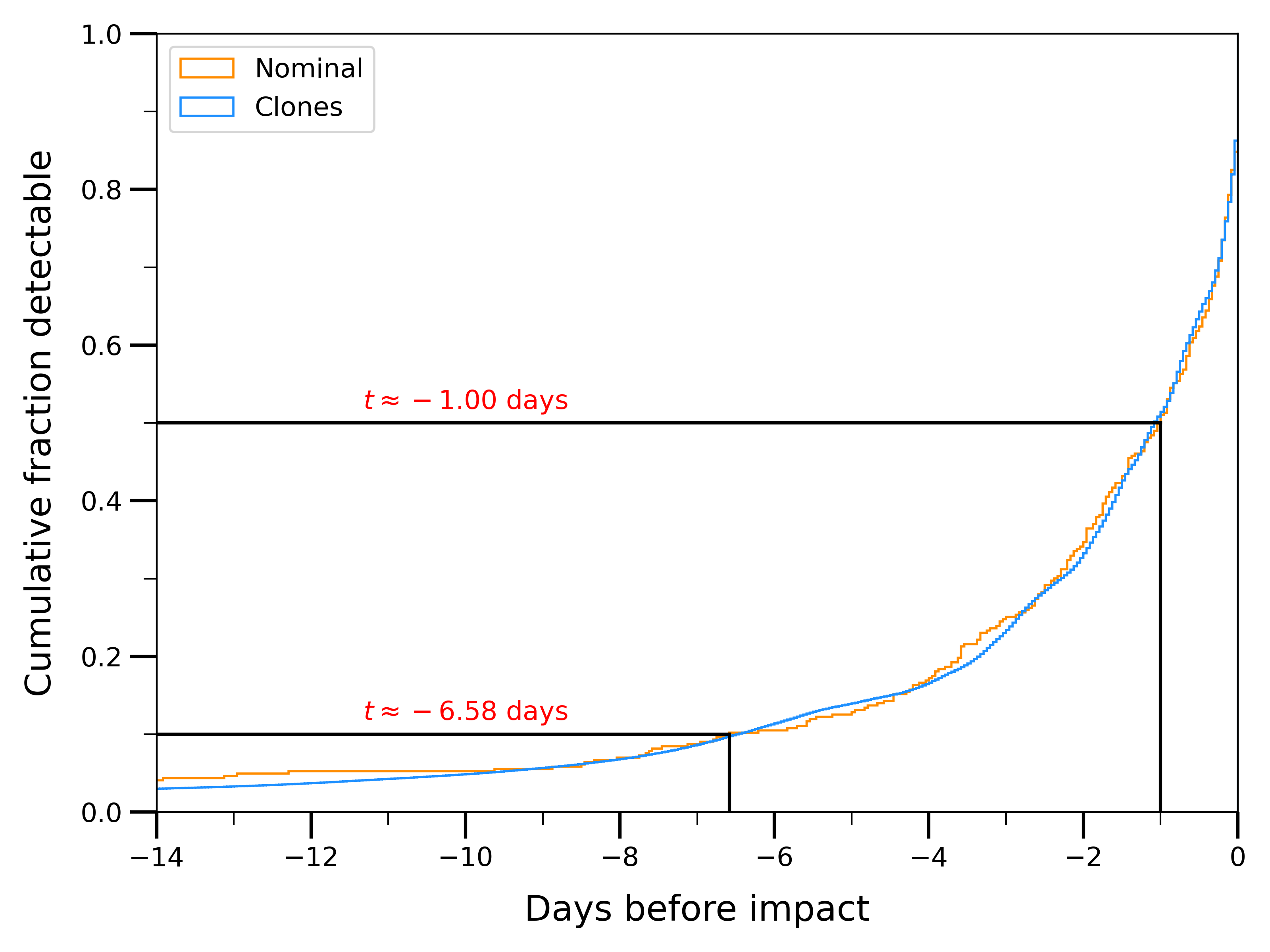}
    \caption{The cumulative fraction of CNEOS-recorded impactors that would become detectable by LSST ($r$-band apparent magnitude $m_r < 24.0$) as a function of time before impact, for nominal orbits (orange line) and when generating $1000$ clones per object (blue line). The black lines indicate the time before impact at which $10\%$ and $50\%$ of nominal CNEOS impactors become visible to LSST.
    }
    \label{fig:cumulative_visible}
\end{figure*}

Overall, the cumulative distribution of the clones appears qualitatively similar to that of the nominal impactors (Figure \ref{fig:cumulative_visible}).
Indeed, comparing the nominal and clone distributions using a two-sample Kolmogorov-Smirnov (K-S) statistical test, we obtain a test statistic of $D \approx 0.03$ and corresponding $p$-value of $p \approx 0.95$. We therefore find no quantitative evidence that the two samples are drawn from different underlying distributions (i.e. the null hypothesis is not rejected) at the $p = 0.05$ confidence level. As such, we suggest that the pre-impact detectability of the CNEOS objects is not significantly affected by orbital uncertainty.


\section{Rubin Impactor Detections: Detailed Simulations} \label{sec:simulated_discoveries}

In this section we perform detailed simulations of LSST's performance to determine the number and properties of CNEOS impactors that would have been observed had LSST been active at the time they impacted. We do not produce clones of each impactor for the analysis carried out in this section, due to computational cost and as the results of Section \ref{sec:visibility} suggest that it would not significantly affect the visibility of the impactors at a population level.

\subsection{Methodology} \label{sec:methodology}

We perform our simulations using \texttt{Sorcha} \citep{merritt_sorcha_2025, holman_sorcha_2025}, an open-source, catalog-level solar system survey simulator specifically designed for large-scale all-sky surveys such as LSST.
\texttt{Sorcha} takes as input a configuration file containing the observatory location, camera field of view, filters, and model of its detection pipeline, a simulated pointing database containing a list of exposures to be taken at given times, and a list of input objects' orbits and physical properties such as colors. It then precisely integrates the objects' orbits using \texttt{ASSIST}, computes their detectability in each exposure (accounting for trailing loss), produces a catalog of astrometric and photometric measurements and uncertainties, and generates a list of simulated observations for each object.
We refer the reader to \citet{merritt_sorcha_2025} for a detailed description of \texttt{Sorcha}.

By default, the LSST pipeline will consider an object to be discovered if it is detected on at least three nights within a span of $15$ days, with at least two observations per night \citep{ivezic_lsst_2019}. 
Here we use \texttt{Sorcha} to simulate discoveries of the CNEOS impactors for LSST, using both the default three-night paradigm and an as-yet unimplemented one-night discovery strategy. In the latter approach, the object's on-sky motion must be sufficiently fast ($>1^\circ$/day) in both exposures to appear as an elongated streak, allowing for the direction and rate of on-sky motion to be constrained. Crucially, this requirement for the object to streak in both exposures essentially eliminates any chance of it being a false positive as the streak lengths, orientations and positions must align precisely for it to be linked.
For all \texttt{Sorcha} simulations in this paper, we use the same \texttt{REBOUND} and \texttt{ASSIST} parameters specified in Section \ref{sec:visibility}. We use the baseline expectations for LSST given by \citet{merritt_sorcha_2025} for all other configuration parameters.

The simulated LSST pointing databases we use are generated using the \texttt{impactor\_startdate} survey strategy of the LSST v5.1 cadence\footnote{\href{https://github.com/lsst-sims/sims_featureScheduler_runs5.1}{github.com/lsst-sims/sims\_featureSchedule\_runs5.1}}, a modified version of LSST's ``baseline" strategy that allows setting an arbitrary start date for the survey. Each pointing database spans $10$ years (the planned runtime of LSST) and together they cover the period from 1994 January 1 to 2026 January 1 (the time range of the impacts considered here).
We use the \texttt{rubin\_sim} software \citep{angeli_end--end_2014, peter_yoachim_lsstrubin_sim_2025} to generate pre-computed sky brightnesses, and the \texttt{rubin\_scheduler} software \citep{naghib_framework_2019, peter_yoachim_lsstrubin_scheduler_2025} to generate sun, moon and planet positions, almanac sunset files, and pre-scheduled ``deep drilling" field (DDF) observations over this time period. 
The reader is referred to \citet{jones_survey_2020} and \citet{schwamb_tuning_2023} for further details regarding LSST's planned cadence and survey strategy. 


We use as input orbits to \texttt{Sorcha} the state vectors computed in Section \ref{sec:visibility}, with each object's reference epoch given by its time of impact.
The colors of each object are then randomly drawn from a multivariate Gaussian distribution fit using kernel density estimation on $384$ S-type or $143$ C-type asteroid spectra (again depending on the object's assumed taxonomic class) from the second phase of the Small Main-belt Asteroid Spectroscopic Survey \citep[SMASS II;][]{bus_phase_2002}, converted to LSST \textit{ugrizy} colours by \citet{kurlander_predictions_2025}.

\subsection{Results} \label{sec:results}

From our \texttt{Sorcha} simulations, we find that $14$ of the $343$ CNEOS impactors were discovered by LSST before impact using the aforementioned discovery criteria. An additional $4$ impactors were not discovered using either criterion but were observed at least once by LSST and as such could theoretically be found in precovery images after impact.

We summarize our simulation results in Table \ref{tab:all_discovered_impactors}, which lists each object's UTC date and time of impact, estimated diameter from CNEOS-reported impact velocity and energy, assumed taxonomic class, method(s) by which the object was found, time of discovery (for objects discovered before impact), earliest observation, total number of observations, and total number of tracklets (sets of at least two observations over a single night).
\begin{sidewaystable*}
    \centering
    \begin{tabular}{l|l|l|l|c|c|c|c}
    \hline
        UTC Impact Date/Time & Diameter & Assumed Class & Method & Time of Discovery & Earliest Observation & Observations & Tracklets \\ 
        (YYYY-MM-DD HH:MM:SS) & (m) & & & (days before impact) & (days before impact) & (total \#) & (total \#) \\ \hline
        2025-09-13 22:24:59 & $2.52$ & S & $1$-night & $1.70$ & $1.72$ & $2$ & $1$ \\
        2023-11-21 22:17:29 & $1.68$ & S & $1$-night & $0.811$ & $0.834$ & $5$ & $1$ \\
        2023-07-26 03:41:53 & $2.68$ & S & $1$-night & $0.812$ & $0.836$ & $2$ & $1$ \\
        2022-07-27 04:41:27 & $3.28$ & C & $1$-night & $0.904$ & $0.933$ & $4$ & $1$ \\
        2022-07-22 00:16:18 & $1.95$ & S & $1$-night & $0.976$ & $0.977$ & $4$ & $1$ \\
        2019-11-28 20:30:53 & $1.84$ & S & $1$-night & $0.795$ & $0.808$ & $2$ & $1$ \\
        2019-05-21 13:12:33 & $5.05$ & C & $1$-night/$3$-night & $5.36$/$6.39$ & $13.4$ & $19$ & $4$/$7$ \\
        2017-06-30 14:26:46 & $2.40$ & S & $1$-night & $1.43$ & $1.46$ & $6$ & $1$ \\
        2015-07-19 07:06:26 & $1.42$ & S & $1$-night & $1.09$ & $2.05$ & $4$ & $1$ \\ 
        2014-06-26 05:54:41 & $2.56$ & S & $1$-night/$3$-night & $23.2$/$8.11$ & $93.0$ & $26$ & $3$/$4$ \\
        2013-04-21 06:23:12 & $4.93$ & S & $3$-night & $9.18$ & $17.1$ & $13$ & $6$ \\
        2012-08-26 14:55:47 & $3.58$ & S & $3$-night & $7.46$ & $18.5$ & $11$ & $3$ \\
        2010-09-03 12:04:58 & $6.45$ & S & $1$-night/$3$-night & $2.43$/$18.2$ & $25.4$ & $24$ & $1$/$9$ \\
        2005-04-19 07:37:47 & $2.89$ & S & $3$-night & $2.13$ & $9.08$ & $10$ & $3$ \\
        \hline
        2025-08-19 14:08:48 & $3.54$ & C & Precovery & & $3.45$ & $3$ & $0$ \\
        2020-10-26 15:07:09 & $2.18$ & C & Precovery & & $0.59$ & $2$ & $0$ \\
        2014-08-23 06:29:41 & $6.39$ & C & Precovery & & $3.93$ & $4$ & $0$ \\
        2005-06-03 08:15:41 & $1.57$ & S & Precovery & & $1.13$ & $2$ & $0$ \\
    \end{tabular}
    \caption{
     All $18$ CNEOS impactors observed by LSST in our \texttt{Sorcha} simulations, including $14$ objects successfully linked and discovered before impact and $4$ that only have precovery observations (were not linked and discovered before impact).
    }
    \label{tab:all_discovered_impactors}
\end{sidewaystable*}
If an object was discovered with both the one-night and three-night methods, the time of discovery and total number of tracklets for both methods are listed. The three-night method can sometimes discover objects earlier than the one-night method, since only exposures where the object is moving faster than $1^\circ$/day can be part of a tracklet in the latter case. The median time of discovery and median earliest observation of the $14$ discovered objects in Table \ref{tab:all_discovered_impactors} are $\sim1.57$ days and $\sim3.06$ days before impact, respectively.

We expect the one-night discovery method to be paramount in allowing LSST to be effective at discovering imminent impactors, as $8$ of the $14$ total impactors discovered in simulations could only have been found using the one-night method. For Rubin to be effective at imminent impactor discovery, it is therefore critical that this discovery method is successfully deployed and not overwhelmed by false positives.



\begin{figure}
    \centering
    \includegraphics[width=1.\linewidth]{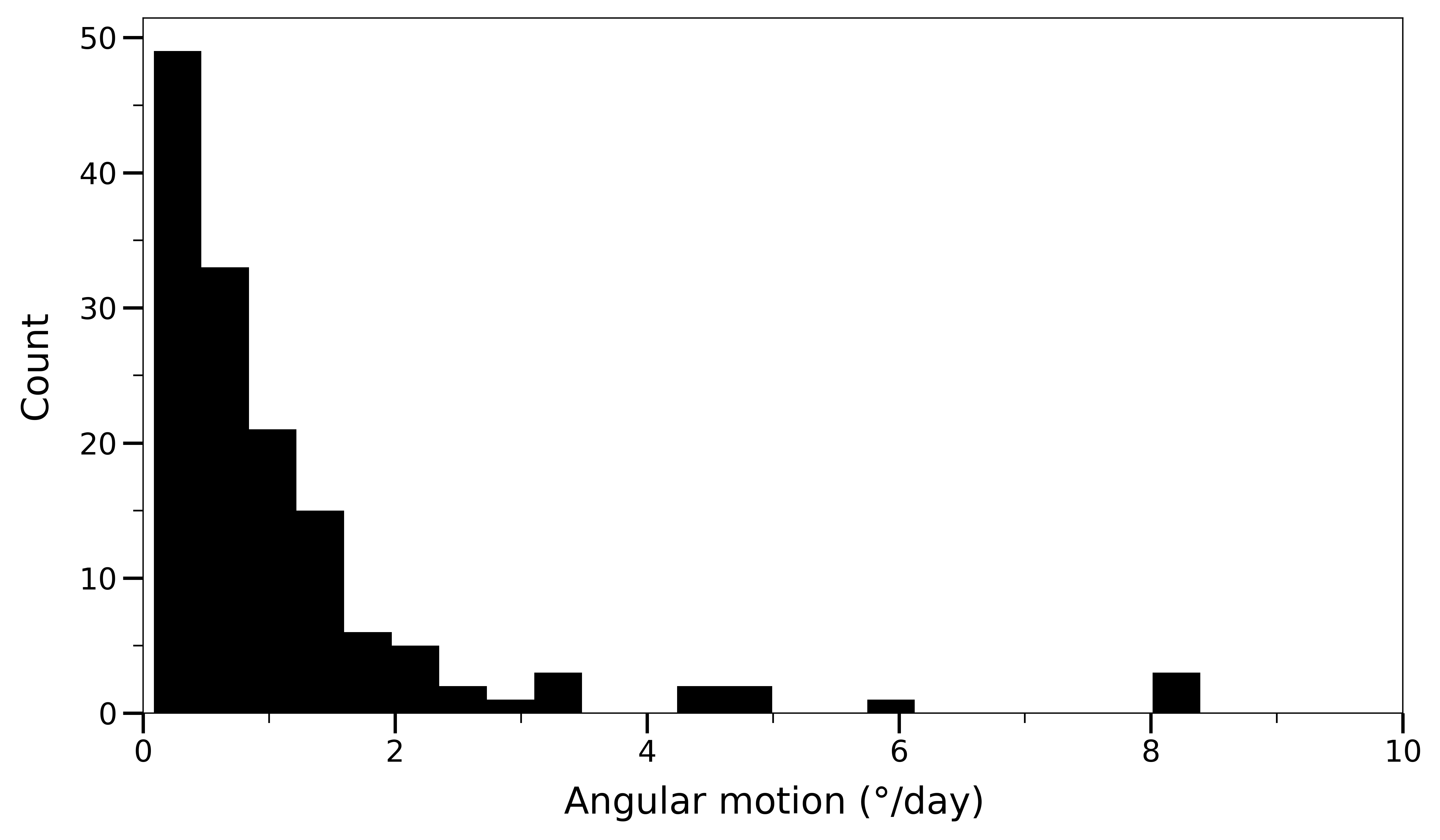}
    \caption{
    The distribution of on-sky angular motions for all observations of the CNEOS impactors in our \texttt{Sorcha} simulations. The angular motions of observed objects are quite small; most move less than $\sim2^\circ$/day, with the largest on-sky motion still less than $10^\circ$. This suggests that LSST observations of imminent impactors are unlikely to be affected by the maximum $10^\circ$/day motion threshold for reporting.
    }
    \label{fig:angular_motion_histogram}
\end{figure}

As imminent impactors are typically fast-moving objects relative to the Earth, we also have to consider their on-sky motion, as LSST imposes a maximum angular motion cutoff of $10^\circ$/day for reporting observations \citep{omullane_dmtn-199_2024}. Figure \ref{fig:angular_motion_histogram} shows the distribution of angular motions for all simulated observations.
Most angular motions are quite small relative to the $10^\circ$/day limit, with even the fastest observed object (imaged $\sim9$ hours before impact) moving at only $\sim8^\circ$/day. We conclude that LSST's $10^\circ$/day motion cutoff is therefore unlikely to affect observations of imminent impactors.

\section{Debiasing the Discovery Yield Estimates}\label{sec:yield_completness}

Our results in Table~\ref{tab:all_discovered_impactors} imply 14 discoveries over the 31-year time span of the CNEOS fireball dataset. Naively assuming that these fireballs are detected at the same rate throughout implies an LSST discovery rate of $1$ imminent impactor every $\sim2$ years. However, the USG sensor network feeding the CNEOS dataset is believed to have become more efficient over time.
Indeed, \citet{pena-asensio_error_2025} note that CNEOS has gradually recorded more high-velocity impactors over time while simultaneously increasing accuracy, with significant improvement from 2018 onward.
While exact details of the USG sensors are classified, \citet{pena-asensio_error_2025} suggest that they may correspond to early warning infrared sensors operated by the U.S. Space Force. \citet{pena-asensio_error_2025} argue that these improvements are a result of the transition from the Defense Support Program to the newer Space Based Infrared System, the latter of which has both greater accuracy and coverage of the Earth. 
This increase is also visible in our own simulations; starting from the first impactor detected in 2005, the detection rate increases throughout the 2010s until reaching $\sim1$/year in the 2020s.
We conclude that this increase in LSST detection rate over time is due to the underlying improvement in completeness of the CNEOS sample and suggest that our discovery rates in later years should be trusted as closer to actual expectations than the early data.

We therefore take the beginning of 2018 as the approximate start date for the deployment of the new, more complete USG sensor system (and therefore a more complete count of the total impactor flux on Earth).
Under this assumption, our simulations from Table \ref{tab:all_discovered_impactors} show that LSST would have discovered $7$ of the CNEOS impactors from 2018 onward, or $0.875$ impactors per year, when using both the one-night and three-night discovery methods. 

\begin{figure}
    \centering
    \includegraphics[width=1.\linewidth]{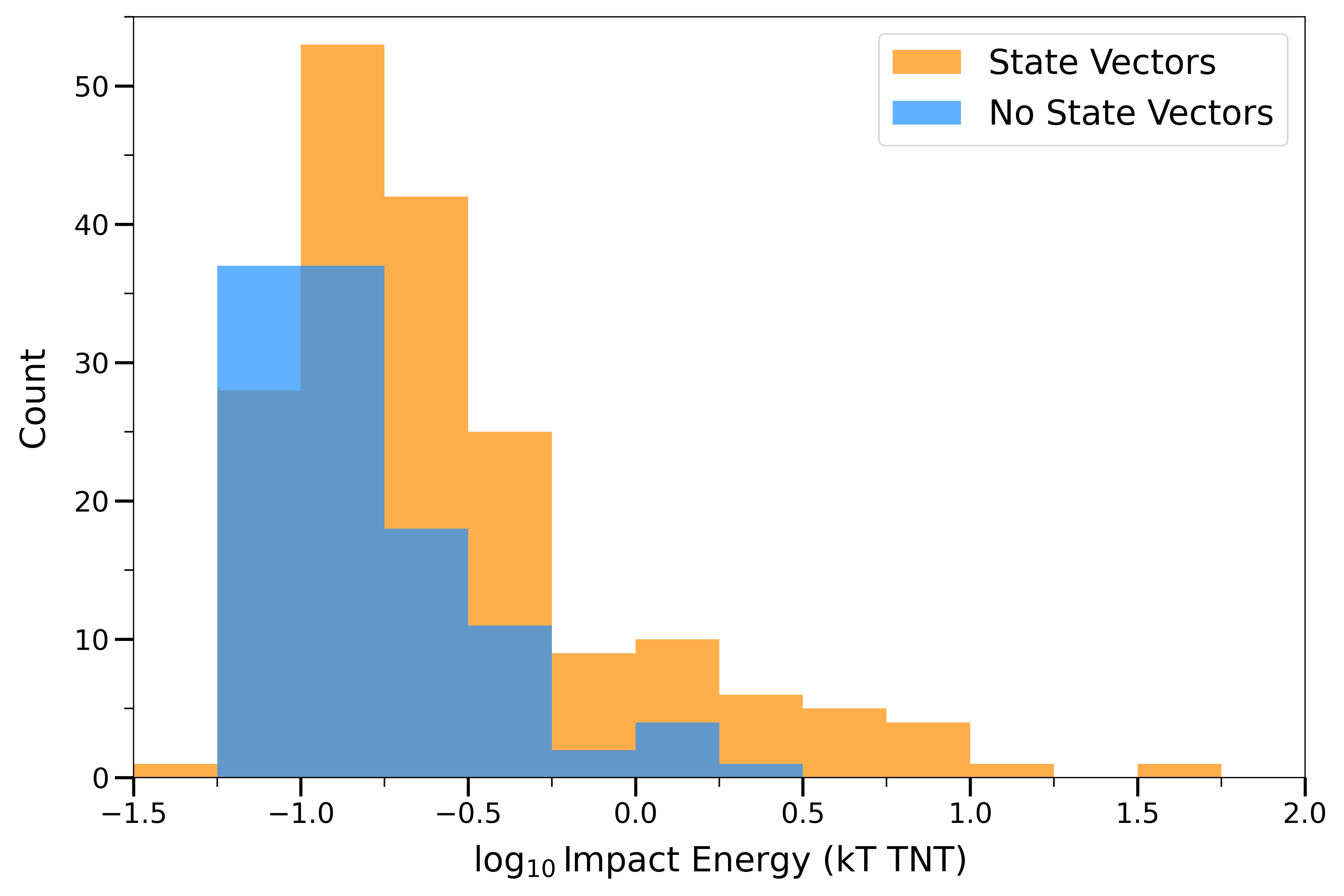}
    \caption{Histograms showing the distribution of $\log_{10}$-impact energies for CNEOS impactors with and without recorded state vectors (orange and blue, respectively), from 2018 January 1 to 2026 January 1. The impact energy distributions appear similar except at the lowest bin.}
    \label{fig:impact_energy_comparison}
\end{figure}

However, our sample only includes impactors with reported state vectors. Of the $295$ fireballs recorded by CNEOS from 2018 January 1 to 2026 January 1, only $185$, or $\sim63\%$, have state vectors. Figure \ref{fig:impact_energy_comparison} shows a comparison of impact energies for those impactors with and without recorded state vectors since 2018 January 1.
The distributions of impact energies for both populations are similar overall, except for the lowest bin.
While the sizes and orbits of those objects without state vectors cannot be directly computed, their impact energies suggest that even accounting for the aforementioned velocity bias, most of these objects would be large enough that they could be seen by LSST before impact.
Extrapolating the rate of $0.875$ impactors discovered per year in simulations to the full CNEOS population, under the assumption that the objects without state vectors could be discovered by LSST at the same rate as those with state vectors, produces an estimated $1.40 \pm 0.14$ discoveries per year, where the errors are the $1\sigma$ uncertainty assuming Earth impacts are modeled as a Poisson process.

We also obtain a similar result by considering the overall fraction of CNEOS impactors discovered by LSST and scaling this value to 
previous estimates of the Earth impactor flux.
In our simulations, $7$ out of $185$ impactors with state vectors since 2018 January 1, or $\sim3.8\%$, were discovered before impact, a fraction which is commensurate with that of the overall sample ($14/343 \approx 4.1\%$).
Scaling this fraction to estimates from fireball and infrasound data that the Earth is impacted by $35-40$ meter-size and larger objects every year \citep[][]{brown_flux_2002, bland_rate_2006, silber_estimate_2009, devillepoix_observation_2019} yields an expected 
$1.14-1.71$ discoveries per year, where the lower and upper bounds are also estimated as the $1\sigma$ Poisson uncertainty on the impact flux range.


Based on these results, we thus expect LSST to discover $\sim1-2$ meter-size and larger imminent impactors per year, comprising $\sim4\%$ of all meter-size Earth impactors, using either the three-night or one-night method.
LSST also discovers $7$ of the $9$ CNEOS impactors observed in our simulations since the start of 2018, and $14$ of $18$ impactors observed overall, suggesting that if both linking methods are implemented, LSST could discover $\sim78\%$ of meter-size impactors that it images.
We note that the overall rate of imminent impactor discovery has significantly accelerated in recent years, with 7 of the 11 known imminent impactors having been discovered since the beginning of 2022 (Table \ref{tab:imminent_impactors}). Taking $1.75$ imminent impactors per year ($7$ objects in $4$ years) as the current rate of discovery means that LSST's estimated $1.40$ imminent impactors per year still represents a near-doubling of the current imminent impactor discovery rate,
especially given the hemisphere bias in impact location for both the known imminent impactors and those impactors observed by LSST in our simulations (discussed in Section \ref{sec:spatial_distribution}).


Finally, we also comment that three of the eleven previously known imminent impactors (2022 WJ1, 2023 CX1 and 2024 BX1) are below the size range analyzed in this work. LSST's higher depth compared to other surveys
means that it will likely observe some of these sub-meter size impactors not considered in our simulations here, suggesting that our estimates should be taken as a lower bound for the total number of imminent impactors that could be discovered.

\section{Discussion}\label{sec:discussion} 

\subsection{Spatial Distribution of Detected Impacts}\label{sec:spatial_distribution}

\begin{figure*}
    \centering
    \includegraphics[width=1.\linewidth]{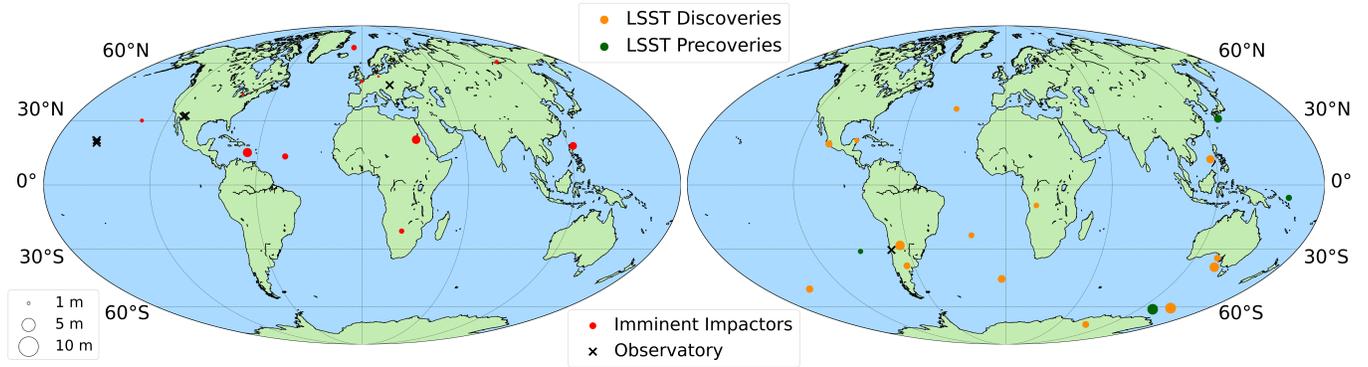}
    \caption{
    A comparison of the impact locations of previously discovered imminent impactors with simulated LSST observations of impactors. The left panel shows the Earth impact locations of the $11$ previous imminent impactors (red dots) and the locations of the observatories that discovered them (black crosses). $10$ of $11$ impacts occurred in the Northern Hemisphere, where all five discovering observatories are also located. The right panel shows the impact locations of all $18$ CNEOS impactors with simulated LSST observations in Table \ref{tab:all_discovered_impactors} (orange/green dots) and the location of the Rubin Observatory in Chile (black cross). In contrast to the known imminent impactors, the majority of impacts ($10$ out of $14$ discovered impactors and $13$ of $18$ overall) occur in the Southern Hemisphere, where Rubin is located.
    The estimated diameter of each object as given in Table \ref{tab:imminent_impactors} (for the known imminent impactors) or Table \ref{tab:all_discovered_impactors} (for the CNEOS impactors observed in our simulations) is indicated by its relative size on the map.
    }
    \label{fig:detection_map}
\end{figure*}

While the spatial distribution of Earth impactors is quite uniform overall (see Figure \ref{fig:cneos_map}), the distributions of both the known imminent impactors and our observed impactors in simulations are appreciably biased.
Figure \ref{fig:detection_map} shows the impact locations of the $11$ known imminent impactors compared to the $18$ CNEOS impactors observed by LSST in our simulations. 
The majority of known imminent impactors ($10$ of $11$) have impacted the Northern Hemisphere, reflecting 
the locations of the five observatories where these objects were discovered (see Table \ref{tab:imminent_impactors}): the Mount Lemmon Survey (part of CSS) and Kitt Peak National Observatory in Arizona, USA, the Konkoly Observatory in Hungary, and ATLAS's Mauna Loa and Haleakal\={a} observatories in Hawai'i, USA, all of which are located in the Northern Hemisphere. Similarly, the majority of impactors observed by LSST in our simulations ($10$ of $14$ discovered objects and $13$ of $18$ observed in total) impact the Southern Hemisphere, where Rubin is located.
These results are consistent with those of \citet{fohring_site_2024}, who found a similar north-south bias when using a set of simulated Earth impactors to estimate LSST's impact on discovery metrics for the proposed Flyeye-2 telescope in different sites.
We therefore expect that Rubin's location in the Southern Hemisphere will nicely complement existing asteroid surveys such as CSS, ATLAS and the Panoramic Survey Telescope and Rapid Response System \citep[Pan-STARRS;][]{kaiser_pan-starrs_2002, kaiser_pan-starrs_2010, chambers_pan-starrs1_2019}, which are primarily based in the north. 
This also means that LSST will observe objects that other asteroid surveys will be unable to detect due to their location, filling a crucial gap in ground-based coverage of Earth impactors.
There does not appear to be a corresponding bias for the Western and Eastern Hemispheres, likely due to the rotation of the Earth.




\subsection{Size Distribution and Albedo of Detected Impactors}

\begin{figure}
    \centering
    \includegraphics[width=1.\linewidth]{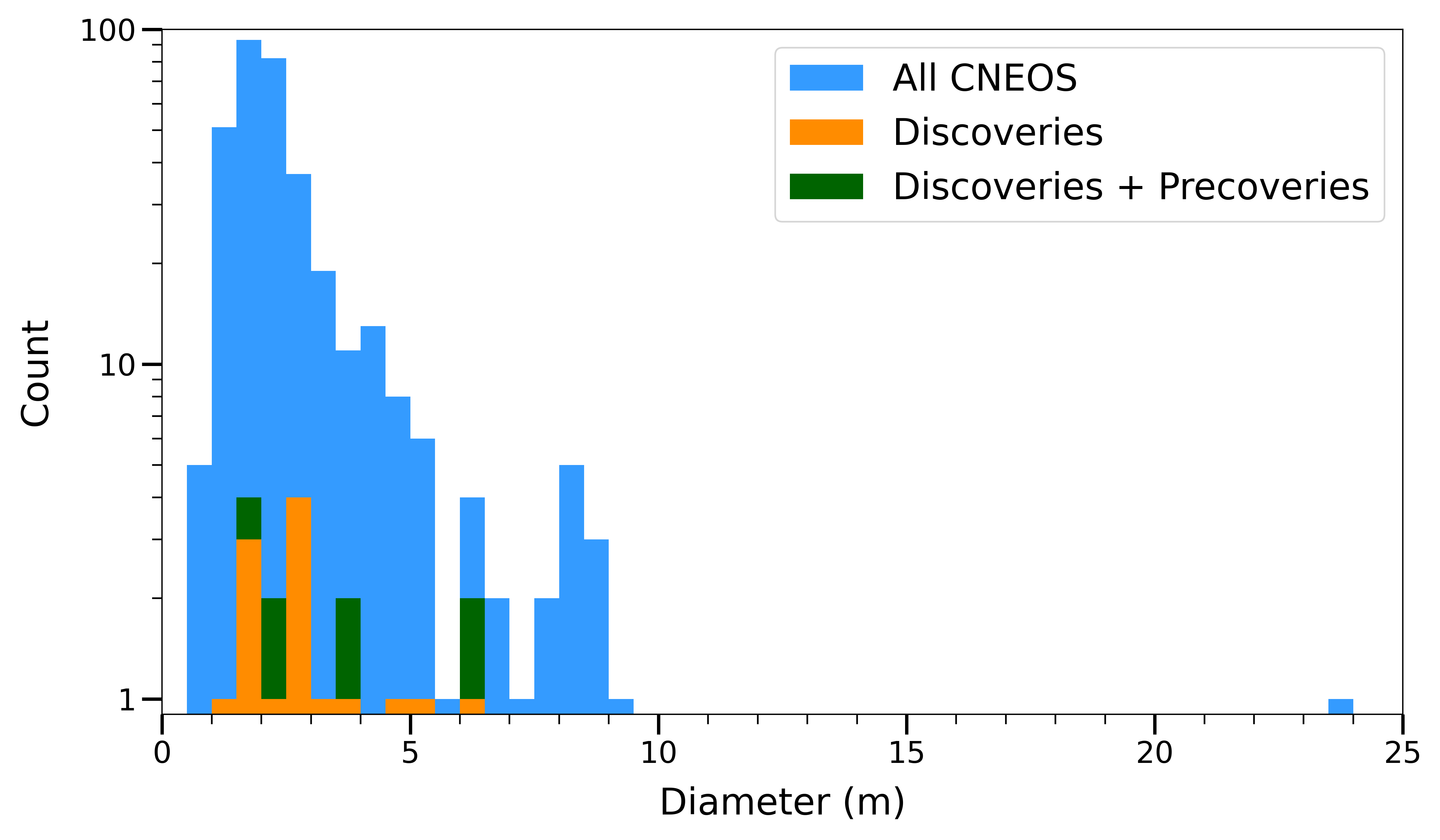}
    \caption{
    The size distribution of the $14$ impactors discovered before impact (orange) and of all $18$ impactors observed at least once (green) by LSST in simulations (Table \ref{tab:all_discovered_impactors}), compared to the overall CNEOS sample (blue). Note that the y-axis is on a logarithmic scale. There is no apparent detection bias with size 
    from $\sim1-10$ m, the approximate range of the dataset. The 2013 Chelyabinsk asteroid (estimated here as $\sim24$ m in diameter) is significantly larger than all other impactors.
    }
    \label{fig:diameter_histogram}
\end{figure}

We examine the size distribution of the observed impactor population in our simulations to search for any detection bias, particularly towards larger objects.
Figure \ref{fig:diameter_histogram} shows that 
the size distribution of observed impactors (both discoveries and precoveries) is similar to that of the overall CNEOS sample, with the lack of detections above $7$ m likely due to Poisson uncertainty.
Comparing the size distribution of the $18$ observed impactors to that of the CNEOS sample as a whole using a two-sample K-S test, we obtain a test statistic of $D \approx 0.29$ and corresponding $p$-value of $p \approx 0.09$, suggesting there is no quantitative evidence that the two samples are from different distributions (the null hypothesis is not rejected) at the $p = 0.05$ confidence level. We thus conclude that there is no apparent observational bias towards larger impactors in the $\sim1-10$ m size regime, roughly the range considered here.





With an expected $\sim1-2$ meter-size and larger imminent impactors discovered every year, it is possible that LSST, in conjunction with other surveys, could observe enough imminent impactors within the duration of the survey to produce the first ever direct telescopic estimates of the size-impact flux distribution of meter-scale Earth impactors.
Such observations could help resolve the aforementioned ``decameter gap" identified by \citet{chow_decameter-sized_2025}.



We also remark that the majority of imminent impactors observed in our simulations ($12$ of the $14$ discovered impactors and $13$ of $18$ observed overall) are S-type asteroids, suggesting there may be a detection bias towards higher albedo objects.
This is likely the result of S-types being brighter overall than C-types, with the albedo difference between equivalent-diameter S- and C-type asteroids corresponding to a change of $\sim1$ magnitude in the size regime considered in this work.

\subsection{New Possibilities for Imminent Impactor Science with LSST }\label{sec:science} 

The median time of discovery in our simulations of $\sim1.57$ days before impact, though short, still represents a longer warning time than for any previously discovered imminent impactor \citep[the longest warning time was for 2014 AA, discovered $\sim21$ hours before impact;][]{farnocchia_trajectory_2016}, with some objects discovered days or even weeks before impact in our simulations.

If these objects could be quickly identified as potential impactors, the additional time for follow-up observations would increase the amount of spectroscopic, high-cadence photometric and potentially polarimetric data from ground-based facilities, allowing albedo, asteroid taxonomy, surface roughness, pole orientation and rotation period to be determined with higher fidelity than has been possible to date.

The longer warning times for imminent impactors in the Rubin era will also enable new observational modes such as radar. Indeed, previous surveys have suggested that some military radars can detect exo-atmospheric meteoroids \citep[][]{Kessler1980}. Modern phased array radar systems, such as EISCAT 3D \citep{Kastinen2020}, can detect meter-size objects out to ranges of $\sim20,000$ km when provided with sufficient warning times (on the order of hours in ideal cases). Such radar information could allow very accurate ranging, refine trajectories and provide information on both rotation periods and surface characteristics at radio wavelengths. 

Longer observational arcs will also improve impactor trajectory precision, making the calculation of meteorite fall locations more accurate and thus facilitating meteorite recovery. For example, \citet{egal_catastrophic_2025} found the respective trajectories determined from astrometric and fireball observations for imminent impactor 2023 CX1 to be within $18$ m of one another. Such high accuracy data allowed meteorites to be found within several hundred meters of predicted fall locations.

For impacts over open ocean, where meteorite recovery is not possible, the additional warning time provided by LSST could make feasible airborne dust sampling of the resulting fireball in lieu of meteorites on the ground, helping provide a compositional ``ground-truth" for a larger sample of imminent impactors \citep{Zolensky1997}.

Finally, some impactors observed but not discovered by LSST may still be found in precovery images if the subsequent fireball trajectory is observed; for example, \citet{clark_preatmospheric_2023} used precovery images from ATLAS of a CNEOS-recorded fireball to constrain the object's trajectory, size, albedo and rotation rate. In such cases, precovery images could still allow much of the aforementioned science to be conducted even if the object is not successfully linked and discovered before impact. 

\section{Conclusions} \label{sec:conclusions}

We model the LSST discovery rate of imminent impactors using $343$ real Earth-impacting asteroids and meteoroids observed globally since 1994 by satellite-based USG sensors as bright meteors in Earth's atmosphere and reported in NASA's CNEOS Fireball and Bolide Database. These objects, ranging from $\sim0.75-24$ m in diameter, represent the largest set of known meter-size and larger Earth impactors to date.

Using the \texttt{Sorcha} Solar System survey simulation software, we generate LSST observations of the CNEOS impactors as if the survey had been running continuously from 1994 to the present day. From these results, we find that LSST could have discovered $14$ of these objects before impact and obtained precovery images for $4$ more.
We use these results to estimate the expected discovery yield and discovery completeness of Earth impactors. Finally, we analyze the properties of the observed impactor population in our simulations for possible detection biases.
Our conclusions are summarized as follows:
\begin{enumerate}
    \item LSST should discover $\sim1-2$ imminent impactors larger than one meter in diameter per year over the course of the survey. This would represent $\sim4\%$ of all meter-size and larger Earth impactors and result in a near-doubling of the current imminent impactor discovery rate since 2022. This estimate assumes Rubin will search for impactors using both the default three-night tracklet linking strategy \citep{heinze_two_2025} and a yet-to-be implemented single-tracklet strategy that would match fast-moving streaks (we adopt $>1^\circ$/day for this threshold) from two exposures taken in a single night. With implementation of this one-night discovery method, LSST could link and discover $\sim78\%$ of all meter-size impactors that it observes. There is no apparent detection bias with size up to $10$ m, the largest objects that routinely impact the Earth. 
    \item LSST's median time of discovery and median time of first observation for impactors discovered in simulations are $\sim1.57$ and $\sim3.06$ days before impact, respectively, and roughly a quarter of discovered impactors were found over a week before impact. This would represent a significantly longer warning time than for previously discovered imminent impactors, all of which were found less than a day before impact.
    \item For Rubin to be an effective discoverer of imminent impactors, it is critical the observatory detects and publicly reports (within hours, and ideally within minutes) tracklets consisting of pairs of trailed observations, as discussed in point 1.
    Relying only on the default three-tracklet strategy would significantly reduce both the number of imminent impactor discoveries and the warning time before impact.
    We do not expect the single-tracklet linking strategy to be adversely affected by LSST's maximum $10^\circ$/day threshold for reporting observations.
    \item We expect LSST to have a detection bias towards objects impacting the Southern Hemisphere, where the Rubin Observatory is located.
    This is consistent with the spatial distribution of previous imminent impactors, which is biased towards impacts in the Northern Hemisphere where the discovering observatories are located.
    LSST's location in the Southern Hemisphere will therefore provide coverage {\em complementary}
    to existing ground-based asteroid surveys primarily based in the Northern Hemisphere, rather than serve to replace them. 
    \item In addition to discovery, LSST will also provide precovery images of some impactors discovered by other surveys or observed as fireballs upon impact. For objects observed only as fireballs, the critical challenge will be to link the state vector at impact to a trailed tracklet of the object in space.
\end{enumerate}

While this paper was in peer review, \citet[][in review]{frazer_simulated_2026} posted a preprint examining the detectability of Earth impactors with LSST under the assumption that objects could be discovered and linked using a single three-observation tracklet (i.e. three observations of an object on the same night). Though this approach will likely be challenging early in the survey due to a significant unknown MBA background and high false positive rate \citep[][]{wagg_expected_2025}, \citet[][in review]{frazer_simulated_2026} show that this detection criterion, on its own, could yield discovery numbers and warning times similar to ours. This further bolsters our claim that the estimates presented here are lower limits: as the survey progresses and the software is improved, the background will become better characterized and more imminent impactors will be detected.

The expected lower limit of $\sim1-2$ imminent impactors discovered per year by LSST will add to the $11$ currently documented imminent impactors, allowing telescopic characterization of Earth impactors at a population level for the first time. Importantly, the near doubling of time-before-impact for most objects, and extension to weeks for some (typically larger) objects, would allow coordinated world-wide observing campaigns for orbit determination, meteor observations, and meteorite recovery of many more imminent impactors to be mounted over the next decade, as well as help inform planetary defense initiatives for larger and more hazardous (but rarer) impactors.




To realize this potential, we hope to implement the single-tracklet discovery method and add it to Rubin's pipelines in 2026. We also encourage the community to begin preparing for rapid reaction follow-up campaigns once Rubin imminent impactor discovery begins.


\begin{acknowledgments}
I.C. sincerely thanks the anonymous reviewer for providing valuable feedback on an earlier version of this manuscript, Eloy Pe\~{n}a-Asensio for providing the original azimuth and altitude angle uncertainties for the calibrated CNEOS fireballs, and Nick Moskovitz and Dave Clark for helpful discussions. 

I.C. acknowledges the support of the Natural Sciences and Engineering Research Council of Canada (NSERC) and is partially funded by an NSERC Postgraduate Scholarship-Doctoral (PGS-D).
The authors also acknowledge the support from the DiRAC Institute in the Department of Astronomy at the University of Washington. The DiRAC Institute is supported through generous gifts from the Charles and Lisa Simonyi Fund for Arts and Sciences, Janet and Lloyd Frink, and the Washington Research Foundation. 

M.J., K.K. and J.M. acknowledge the support of the Asteroid Institute, a program of B612 Foundation, made possible by leadership gifts to the ADAM project from the W. K. Bowes Jr. Foundation, the McGregor Girand Charitable Endowment, the P. Rawls Family Fund, Tito’s CHEERS, Alison and Steve Krausz, the Lyda Hill Foundation, Maryann and John Montrym, Google Cloud, and three anonymous donors. I.C. is grateful to Colleen Fiaschetti, Alec Koumjian, Ed Lu, Danica Remy, Nate Tellis, and Delphine Veronese-Milin for providing valuable feedback and supporting the author at the December 2025 Asteroid Institute coding sprint.

This manuscript is based in part on work supported by the National Aeronautics and Space Administration (NASA) under Grant No. 80NSSC24K0852 issued through the Yearly Opportunities for Research in Planetary Defense (YORPD) Program.
Additional funding for this work was provided by the Meteoroid Environment Office of NASA through co-operative agreement 80NSSC24M0060.
This research has made use of the Astrophysics Data System (ADS), funded by NASA through co-operative agreement 80NSSC21M0056.
\end{acknowledgments}

\software{\texttt{REBOUND} \citep{rein_rebound_2012}, \texttt{ASSIST} \citep{holman_assist_2023, rein_matthewholmanassist_2023}, \texttt{Sorcha} \citep{merritt_sorcha_2025, holman_sorcha_2025}, \texttt{rubin\_sim} \citep{angeli_end--end_2014, peter_yoachim_lsstrubin_sim_2025}, \texttt{rubin\_scheduler} \citep{naghib_framework_2019, peter_yoachim_lsstrubin_scheduler_2025}.
}

\bibliography{Imminent_impactors}{}
\bibliographystyle{aasjournalv7}



\end{document}